\documentclass[sigconf, dvipsnames]{acmart}

\copyrightyear{2020} 
\acmYear{2020} 
\setcopyright{acmcopyright}
\acmConference[SIGIR '20]{Proceedings of the 43rd International ACM SIGIR Conference on Research and Development in Information Retrieval}{July 25--30, 2020}{Virtual Event, China}
\acmBooktitle{Proceedings of the 43rd International ACM SIGIR Conference on Research and Development in Information Retrieval (SIGIR '20), July 25--30, 2020, Virtual Event, China}
\acmPrice{15.00}
\acmDOI{10.1145/3397271.3401095}
\acmISBN{978-1-4503-8016-4/20/07}

\usepackage{graphicx}
\usepackage{amsmath}
\usepackage{array}
\usepackage{booktabs}

\usepackage{multirow}

\usepackage{bbold}
\usepackage{mathtools}
\usepackage{xcolor}
\usepackage{arydshln}

\newcolumntype{C}[1]{>{\centering\arraybackslash}m{#1}}

\definecolor{colour1}{RGB}{204,0,0}
\definecolor{colour2}{RGB}{255,102,102}
\definecolor{colour3}{RGB}{204,153,0}

\begin{document}
\fancyhead{}

\title{Learning Colour Representations of Search Queries}

\author{Paridhi Maheshwari}
\affiliation{\institution{Adobe Research}}
\email{parimahe@adobe.com}

\author{Manoj Ghuhan}
\affiliation{\institution{Adobe Research}}
\email{ghuhana@adobe.com}

\author{Vishwa Vinay}
\affiliation{\institution{Adobe Research}}
\email{vinay@adobe.com}

\renewcommand{\shortauthors}{Maheshwari et al.}

\newcommand\blfootnote[1]{
  \begingroup
  \renewcommand\thefootnote{}\footnote{#1}
  \addtocounter{footnote}{-1}
  \endgroup
}

\begin{abstract}
Image search engines rely on appropriately designed ranking features that capture various aspects of the content semantics as well as the historic popularity. In this work, we consider the role of colour in this relevance matching process. Our work is motivated by the observation that a significant fraction of user queries have an inherent colour associated with them. While some queries contain explicit colour mentions (such as `black car' and `yellow daisies'), other queries have implicit notions of colour (such as `sky' and `grass'). Furthermore, grounding queries in colour is not a mapping to a single colour, but a distribution in colour space. For instance, a search for `trees' tends to have a bimodal distribution around the colours green and brown. We leverage historical clickthrough data to produce a \textit{colour representation} for search queries and propose a recurrent neural network architecture to encode unseen queries into colour space. We also show how this embedding can be learnt alongside a cross-modal relevance ranker from impression logs where a subset of the result images were clicked. We demonstrate that the use of a query-image colour distance feature leads to an improvement in the ranker performance as measured by users' preferences of clicked versus skipped images.
\end{abstract}



\maketitle

\section{Introduction}
A successful search relies on the ability to accurately interpret the intent behind a user's query and placing the relevant items high up in the ranked results. In this work, we are interested in the search for images via textual queries. Recent work on this application falls into two main buckets: (i) developing evaluation methods and metrics that capture the specifics of this scenario~\cite{zhang2018well, xie2018constructing}, and (ii) vertical-specific aspects of cross-modal representation learning~\cite{laenen2017cross, carvalho2018}.

We deal with historical clickthrough data from a commercial image-search engine, where the items in the index are professionally taken high-quality images to be used for publishing purposes. The users of such a system exhibit similar properties to that of a general-purpose search engine (e.g. web) -- the use of relatively short queries with multiple reformulations within a session. The notion of colour is central to this scenario, reflected by the fact that user queries very often contain colour words. These words typically act as adjective qualifiers for objects (such as ``red balloons'' or ``yellow brick road'') or occur in coordinated pairs (such as ``red and blue background'' and ``black and white dog'').

Various aspects of colour - its naming, perception, affect and effective use - have been very well-studied and are detailed in the next section. In the current work, we utilise clickthrough data captured by a functioning search system to map queries (and their constituent terms) to a representation of colour. In simple terms, we map the query ``yellow brick road'' into a colour space using historic click logs pertaining to the query. Leveraging behavioural data of users in this manner allows for a data-driven alternative to a well-studied subject in psycholinguistics and computer vision. While interesting in its own right, we believe that a better understanding of the colour properties of queries will help improve the ranking quality of future searches and we support this hypothesis with empirical results in Section~\ref{seq:colourFtr}.\blfootnote{*to be read in colour print}

The richness and ambiguities of language make this a challenging task. For example, while the term `black' in ``black beach Santorini'' provides some indications of the colour profile of the desired result, it is unlikely that the same term had any literal significance in the query ``Black Friday''. Building accurate models for colour would require the identification of such false positives. Similarly, consider the query ``black coffee in white cup''. While detailed image analysis might allow us to localise the main concepts (`coffee' and `cup'), it will require a robust understanding of the colour intent to effectively address the query (e.g. demoting pictures of black coffee in cups of other colours). We take initial steps in this direction by considering the novel task of building a text to colour encoder and demonstrating how this colour feature can improve search ranking.

Datasets mapping colour words to points in colour space exist, and we extend these in two ways: (a) mapping a text phrase to a distribution in colour space by capturing the range of associated images in historical interactions for a given query, and (b) learning representations for phrases with implicit notions of colour, e.g, to build an understanding that ``concrete walls'' are grey even if the colour word dictionary does not contain the term `concrete'.

To summarize, our main contributions are as follows:
\begin{enumerate}
    \item We propose a colour representation for search queries by incorporating user preferences from the historical clickthrough data and subsequently learn a query $\rightarrow$ colour encoder.
    \item We demonstrate that the use of colour representations of the user query and the image as a ranking feature results in improved performance of the search engine.
    \item We also propose an end-to-end framework to jointly learn the colour representation of queries while simultaneously training the ranker for the search engine.
\end{enumerate}

\section{Related Work}
In this section, we detail out the related work, focusing on the central task of mapping text units (i.e., words or phrases) to colour. 

\subsection{Colour Names and Cognition}
The understanding of colour names and their effect on human cognition has long been studied by psychologists. The constitutional work by Berlin and Kay~\cite{berlin1991basic} and its extension to the World Colour Survey~\cite{kay2009world} ascertained the shared existence of $11$ basic colour names in English and several other European languages. Moreover, experiments with different linguistic categories of colour concluded that differences in colour naming lead to differences in colour discrimination~\cite{roberson2000color, winawer2007russian}.

The fundamental link between colour and language is demonstrated by the Stroop effect~\cite{stroop} wherein conflicting textual and colour indications increase the cognitive load and cause interference. Subjects take longer to distinguish colours when the colour of the ink and the colour described by the text do not match. A similar effect was observed for words with strong colour associations such as {\color{red}fire} and {\color{Green}grass}~\cite{stroop2}. This interference is evoked by differences in the colours perceived from the language and the colour of the text. In contrast to this phenomenon, colours which are semantically resonant with the text, also known as memory colours, facilitate cognition. The association between language and colour is eminent when colour is an important aspect of the concept it refers to (for example, {\color{Dandelion}banana} and {\color{Cerulean}sky})~\cite{conklin1955hanunoo}. Lin et al.~\cite{lin2013selecting} further validated that such a colour assignment to represent graphs improves the speed of reading tasks by enhancing the memorability of chart legends.

\subsection{Modelling Colour Associations}
One of the early approaches to bridge the gap between colour and language involved fitting statistical models to human judgements of colour-name associations~\cite{menegaz2006discrete, lindner2012large}. These works curate datasets by asking subjects to map a colour patch to the $11$ basic colours. Following this, extensive colour-naming surveys were set up by Nathan Moroney~\cite{moroney2003unconstrained} and Randall Monroe (known as XKCD)~\cite{xkcd} to overcome the constraint of fixed vocabularies. These large datasets provide a means for data-driven modelling approaches. Heer et al.~\cite{heer2012color} employ a probabilistic model to encode relationships in colour naming datasets. Other works leverage topic modelling techniques such as Probabilistic Latent Semantic Analysis~\cite{van2009learning} and supervised Latent Dirichlet Allocation~\cite{schauerte2012learning} to learn these associations. Owing to the success of machine learning, more recent works propose neural network architectures to predict points in colour space given their names~\cite{kawakami2016character, monroe2017colors}.

A closely related task is the inverse problem of mapping colours to their names. McMahan \& Stone~\cite{mcmahan2015bayesian} and Meo et al.~\cite{meo2014generating} present Bayesian colour naming models, while Monroe et al.~\cite{monroe2016learning} use recurrent neural networks to predict a sequence of colour terms iteratively, similar to conditional language models. A common limitation of the aforementioned methods and datasets is the skewed nature of their vocabularies - predominantly containing colour descriptors (e.g., bluish green, light pink) rather than commonplace objects that have a strong colour intent. The focus of our work is to map arbitrary word sequences into colour space by utilizing behavioural data of users.

In the closest related work, Hasavi et al.~\cite{havasi2010automated} use a crowdsourced knowledge base called ConceptNet to form colour-word associations. For unseen words or phrases, they identify a colour by interpolating between known words, which are semantically similar to the phrase. Another line of work involves the use of image search websites to learn the colour mapping. Lindner et al.~\cite{lindner2012color} propose a statistical framework on relevant images from Flickr to determine the associated colour for a concept. They also compute a colourability score which determines the validity of the colour assignment for the expression. This work was further extended to create semantic colour palettes~\cite{lindner2013automatic}. However, these algorithms return a single best colour and do not allow for multiple valid colour options (e.g., apples can be both red and green).

To address this problem, Lin et al.~\cite{lin2013selecting} introduce an algorithm to automatically determine label-colour affinity scores by analyzing images from a Google search and then generate an optimal matching between chart values and a target colour palette. The work of Setlur and Stone~\cite{setlur2015linguistic} expands on their work by refining the queries using additional semantic information to define the context. These techniques do not learn from existing data and rely on Google image search to represent a concept in colour space.

To this end, we leverage machine learning models and propose a data-driven technique to learn the mapping from language (as represented by search queries) to colour. Additionally, unlike prior work, we do not restrict ourselves to a controlled vocabulary.

\section{Background}
Our objective is to build models that map an input sequence of words representing a search query into a colour embedding. In Section~\ref{sec:colourSpaces}, we introduce one possible representation of colour by quantising an existing colour space. While the experiments described in the current paper utilise this specific representation, this is not a constraint of our work. Given images under some chosen colour representation, our proposed model aims to embed queries into the same colour space. We further argue that colour plays an important role in an image-search setting. To do this, we need a technique to measure distances between the colour representations of the query and items, and this is detailed in Section~\ref{sec:colourSimilarity}.

\subsection{Colour Spaces} \label{sec:colourSpaces}
Though physically rooted in the perception of light of different wavelengths, there are multiple mathematical models for colour depending upon the application setting \cite{wyszecki1982color}. The widespread consumption of images on electronic displays engendered the Red-Green-Blue (RGB) family of colour spaces. This is an additive model where every displayable colour is a weighted combination of the $3$ primary colours and an individual pixel is represented as a point in $3-$dimensional space. Each axis can take an integer value between $0$ and $255$, implying that a total of $256*256*256$ unique colours can be represented by an RGB tuple. Distance between any two colours can be computed as the Euclidean distance between two locations in the $3$D RGB space.

For user-facing applications, it is of interest to design distance functions between colours that accurately capture the colour differences as perceived by humans. A method to enable this would be to non-uniformly transform the RGB space such that perceptually similar shades are placed together geometrically. One such colour space is the Hue-Chroma-Luminance (HCL) model~\cite{sarifuddin2005new}, where each pixel is again a point in a $3-$dimensional space. In Figure~\ref{RGBversusHCL}, two candidate colours are chosen such that they are at the same Euclidean distance in RGB space with respect to a reference colour (top row). The corresponding distances in HCL space, however, more accurately reflect the fact that the colour in the middle row \textcolor{colour2}{\rule{0.22cm}{0.22cm}} is \textit{visually more similar} to the reference colour \textcolor{colour1}{\rule{0.22cm}{0.22cm}} than the colour in the last row \textcolor{colour3}{\rule{0.22cm}{0.22cm}}. The HCL model, apart from having colours distributed in a more perceptually uniform manner, has an additional advantage that $3$ dimensions are more intuitive to a human observer. 

\begin{table}[!ht]
\caption{Advantage of perceptually uniform colour spaces.}
\label{RGBversusHCL}
\centering
\begin{tabular}{ c | cc }
    \raisebox{-0.45\height}{\includegraphics[width=4mm]{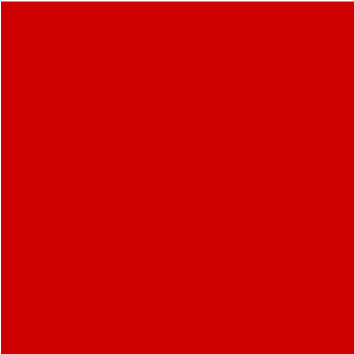}} & RGB & HCL \\[1ex]
    \midrule
    \raisebox{-0.45\height}{\includegraphics[width=4mm]{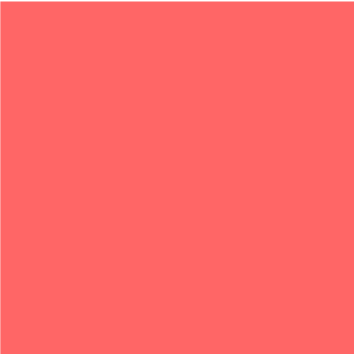}} & $153$ & $34.6$ \\[1.5ex]
    \raisebox{-0.45\height}{\includegraphics[width=4mm]{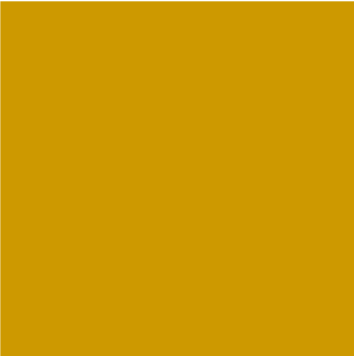}} & $153$ & $65.4$ \\
\end{tabular}
\end{table}

The International Commission on Illumination (CIE) has defined the \textit{LUV} and \textit{LAB} colour spaces, with the `$L$' corresponding to lightness/luminance and the other two channels representing chrominance. The perceived difference between colours is proportional to the distance in these colour spaces. There are well known conversions from one coordinate system into the others, with the application setting typically dictating the choice of colour space.

\begin{figure}[!ht]
\centering
\includegraphics[width=\columnwidth]{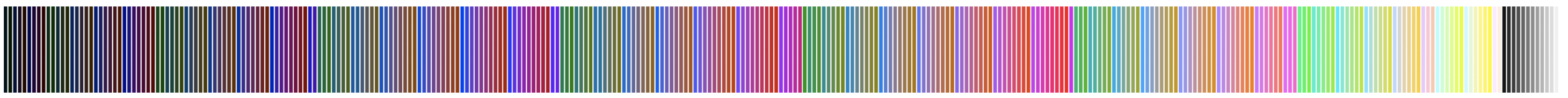}
\caption{Colour space quantised into B = 327 points.}
\label{colourBins}
\end{figure}

Given a choice of colour space, the colour representation of an image is taken to be its distribution over all possible colour tuples. Rather than utilise the entire fidelity, it is a common practice to discretise the colour space and use only a set of representative points~\cite{smith1996visualseek}. Owing to the perceptual uniformity of the HCL space, uniformly dividing it into $B$ points would span the range of colours as perceived by humans. This quantisation, however, involves a trade-off: a smaller $B$ leads to an increased loss in information but a smaller size for an index maintaining colour information for a set of images. For the experiments described in this paper, $B=327$ has been chosen as the operating point and the resulting colour points are shown in Figure~\ref{colourBins}.

To obtain the colour representation of an image, each pixel is assigned to the nearest point among the predefined $B$ colours. This histogram is then normalised to sum to $1$, resulting in a distribution over the bins. We define this as the \textit{colour representation} of the image and it indicates what fraction of the image maps to the colour of a particular bin. To illustrate the qualitative performance of our model in later sections, we provide visualisations for this colour embedding, where the height of the bar represents the weight of that bin, and the colour of the bar corresponds to its HCL value.

\begin{figure}[!ht]
\centering
\includegraphics[width=0.95\columnwidth]{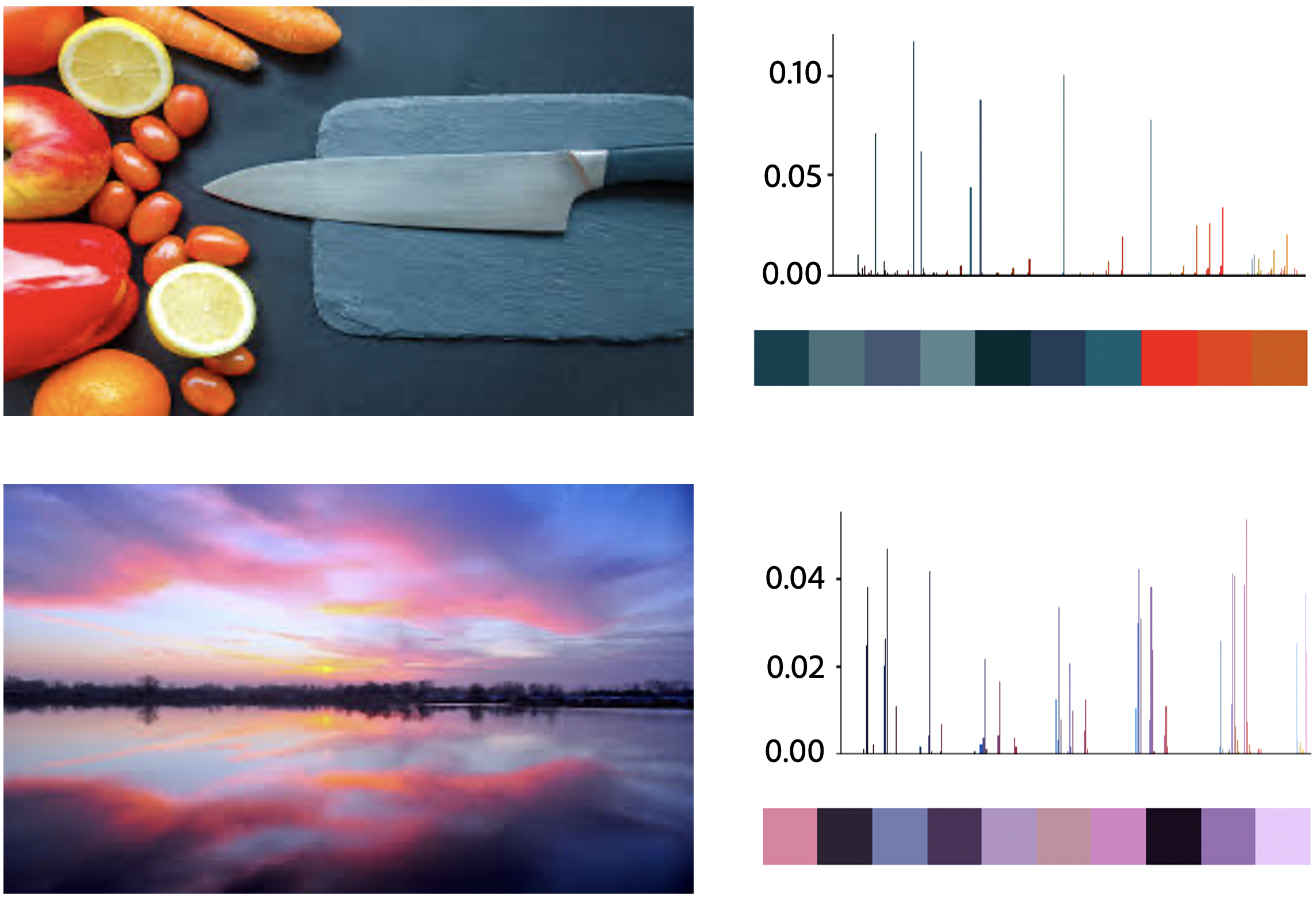}
\caption{Example images, their colour representations and top 10 bins of the histogram. The first image predominantly contains shades of grey and orange, whereas the second image has a mix of hues of pink and blue.}
\label{imageAndColourRep}
\end{figure}

\subsection{Colour Similarity and Retrieval} \label{sec:colourSimilarity}
In Content-Based Information Retrieval (CBIR), the retrieval system invokes a ranking function that computes the relevance between the query and candidate items. For image-centric CBIR, where the query is also an image, the ranking function incorporates notions of similarities between images. Colour is typically a dominant feature in such a ranking function~\cite{cox2000bayesian}, therefore requiring a similarity measure between the colour representations of the query image and candidate results. Note that computing the \textit{colour distance} between two images is different from computing the distance between individual pixels represented in a colour space (the previous section). 

The application scenario considered here is traditional cross-modal retrieval -- we convert the text query into its colour representation and utilise its distance from candidate images' colour representations as one factor amongst others in a relevance ranking function. We provide three options here, which we evaluate in our experimental section as objectives for training the query $\rightarrow$ colour encoder model.

\begin{enumerate}
    \item \textbf{Kullback-Leibler Divergence}: This function treats the colour histograms as discrete probability distributions and measures the difference as
    \begin{displaymath}
        \mathrm{D_{KL}} (P, Q) = \sum_i^B P_i \log \bigg( \frac{Q_i}{P_i} \bigg)
    \end{displaymath}
    \item \textbf{Histogram Intersection}: We compute similarity as the extent of overlap between the histograms and define the loss as its additive inverse. To ensure a positive loss, we add the maximum value of similarity (i.e., $1$).
    \begin{displaymath}
        \mathrm{D_{HI}} (P, Q) = 1 - \sum_i^B \min (P_i, Q_i)
    \end{displaymath}
    \item \textbf{LUV Distance}: We follow the authors of \cite{lee2016automatic} and represent the colour histogram using statistics from the chrominance and luminance channels separately. This is computed by utilising well-known conversions between HCL and LUV. The distance is defined as:
    \begin{displaymath}
        \mathrm{D_{LUV}} (P, Q) = {\mathcal{D}_e(L_P, L_Q)}^{2} \cdot {\mathcal{D}_h(\mathcal{N}_P, \mathcal{N}_Q)}^{1}
    \end{displaymath}
    where $\mathcal{D}_e$ represents the Euclidean distance between the luminance features $L_P$ \& $L_Q$.  $\mathcal{D}_h$ is the Hellinger distance, which for the specific case of multivariate Gaussians is computed as
    \begin{displaymath}
        \mathcal{D}_h(\mathcal{N}_P, \mathcal{N}_Q) = 1 - \dfrac{|\mathrm{\Sigma}_P\mathrm{\Sigma}_Q|^{1/4}}{|\bar{\mathrm{\Sigma}}|^{1/2}} \exp \bigg(-\frac{1}{8} \bar{\mu}^{T} \mathrm{\bar{\Sigma}}^{-1} \bar{\mu} \bigg)
    \end{displaymath}
    \begin{displaymath}
        \text{where} \,\,\,\, \bar{\mu} = |\mu_P - \mu_Q| + \epsilon \text{   and   } \bar{\mathrm{\Sigma}} = \frac{\mathrm{\Sigma}_P + \mathrm{\Sigma}_Q}{2}
    \end{displaymath}
    $\mathcal{N}_P = (\mu_P, \mathrm{\Sigma}_P)$ \& $\mathcal{N}_Q = (\mu_Q, \mathrm{\Sigma}_Q)$ are the summary statistics for the $2D$ Gaussian chrominance channels of $P$ and $Q$, and $\epsilon=1$. This option has an additional modelling assumption -- that of using a single Gaussian for chrominance.
\end{enumerate}

Similar to colour spaces, notions of colour differences are also well studied. While we propose 3 alternatives here, other combinations of colour space + distance functions are possible, e.g. the CIEDE2000 metric. Our primary objective remains that of learning a colour representation of queries as distributions over the $B=327$ bins as described in Figure~\ref{colourLabels}. We leave the exploration of other colour spaces and distance functions, as well as their impact on the learning ability of query $\rightarrow$ colour models, as future work.

\section{Learning Colour Representations} \label{learningColour}
Our main data is the logged impression data from Adobe Stock\footnote{\url{https://stock.adobe.com/}}. The user provides a query as input, and in response, relevant images are retrieved and displayed in a grid. For every user query, the impression log contains information about the images in the result set, their positions in the ranked list, and the subset of images which were clicked by the user. Some user clicks on the result page lead to a downstream conversion, i.e., the images are licensed for future use. The grades of implicit signals available from the logs are indicators of relevance to different degrees. That is to say, while a purchase is strong evidence for query-to-image relevance, the use of the noisy but plentiful clickthrough data is now commonplace.

In order to capture feedback from user behaviour effectively, we filter queries with a minimum of $10$ displayed and $4$ clicked images. We further constrain the queries to contain a maximum of $6$ words as only $1\%$ queries have word count greater than $6$. This results in a total of $15,627$ queries spanning $457,156$ unique images. Additionally, every image contains the following metadata - tags describing the content and a caption provided by the image creator. The statistics of the final dataset are summarized in Table~\ref{datasetStatistics}.

\begin{figure}[!ht]
\centering
\includegraphics[width=\linewidth]{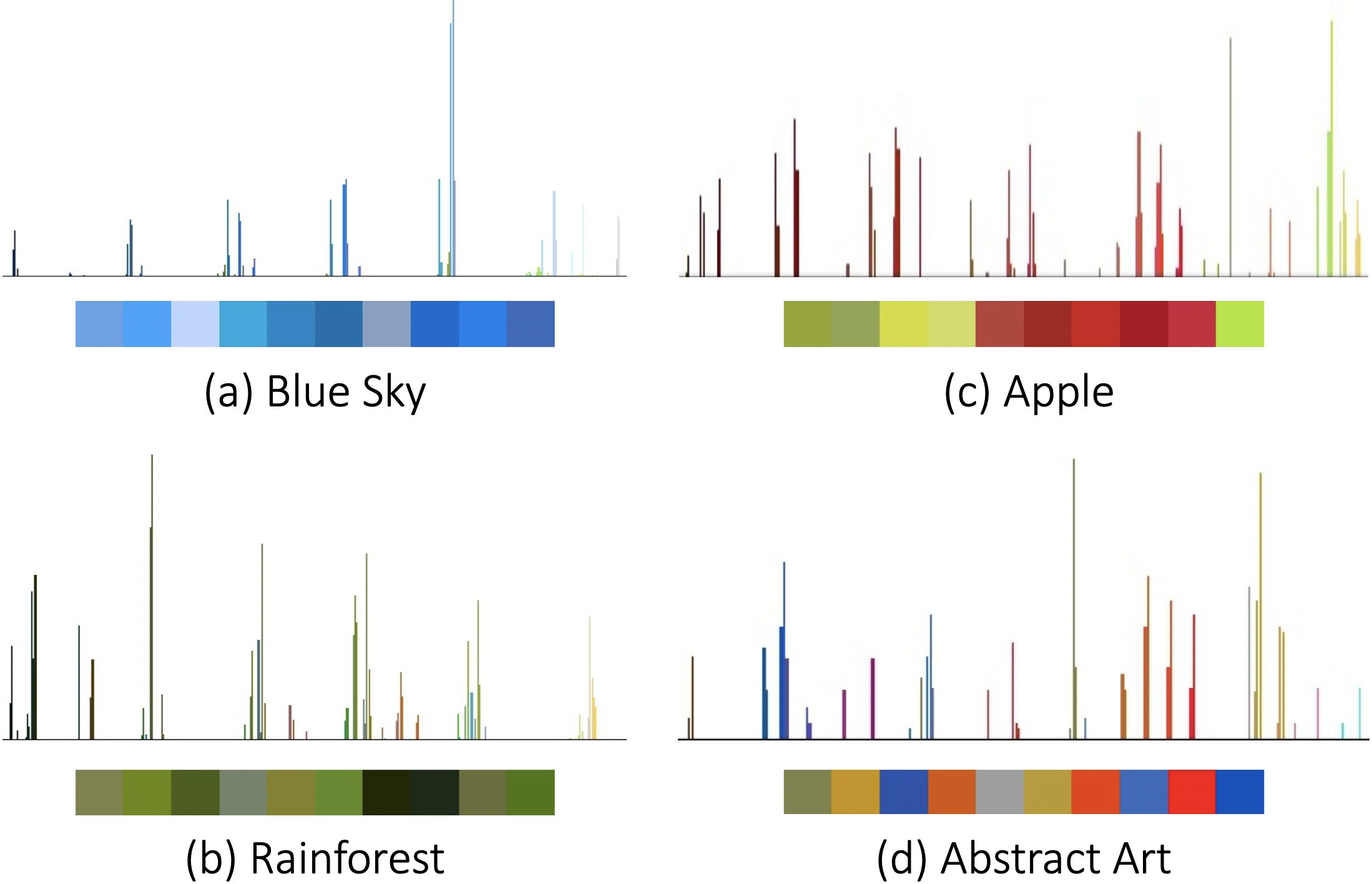}
\caption{Sample queries, their ground-truth colour representations and the top 10 bins shown as a palette. The queries have the following colour intents: (a) explicit mention (b) implicit (c) implicit bi-modal (d) uniform. Note that we interpret a flat distribution over colour bins as an absence of colour intent.}
\label{colourLabels}
\end{figure}

The target colour representation for a query is defined as the average of the colour histograms of the clicked images for that query. To put it mathematically, a query $q$ with $m$ displayed images given by $\mathrm{I_q} = \{I_q^1,\,I_q^2\,..\,I_q^m\}$ and clicked-or-not boolean variable $\mathrm{r_q} = \{r_q^1,\,r_q^2\,..\,r_q^m\}$ is embedded in colour space as,
\begin{displaymath}
    C_q = \frac{1}{m_c} \sum_{i=1}^m r_q^i * C_i \,\,\,\, \text{where} \,\,\,\, m_c = \sum_{i=1}^m r_q^i
\end{displaymath}
Here, $C_i$ is the histogram corresponding to image $I_q^i$ and $m_c$ is the number of clicked images for query $q$. This mapping is illustrated in Figure~\ref{colourLabels} for some queries with inherent colour intent.

Such an average can be quite noisy -- e.g, for the query ``red rose'', different users might prefer images with varying proportions of the image covered by the red rose itself, as well as any surrounding artifacts like green leaves. Therefore, the average across clicked images might lead to an aggregate distribution with peaks at both red and green regions of the colour space. In experiments described later, we provide evidence that the data contains enough signal to bring out the expected modes in the colour representation. 

\begin{table}[t]
\caption{Statistics of the query log data from Adobe Stock.}
\label{datasetStatistics}
\centering
\begin{tabular}{r C{1.75cm} }
    \toprule
    \# queries & $15,627$ \\
    \# displayed images per query & $81.57$\\
    \# clicked images per query & $9.66$\\
    \midrule
    \# images & $457,156$ \\
    \# tags per image & $39.89$ \\
    \# words in caption per image & $6.06$ \\
    \bottomrule
\end{tabular}
\end{table}

Note that for the purposes of training the query $\rightarrow$ colour encoder model, we could have used the average colour histogram of images in the impressions as ground-truth, as in psuedo-relevance feedback. This alternative would lead to a model that represents what the engine already knows implicitly, while our objective is to factor in users' preferences. In Section~\ref{multitask}, by contrasting the clicked images with those shown but not clicked by users, we more accurately interpret user preferences and utilise this signal towards our objective of building models that translate a textual query into a colour embedding.

\subsection{Query $\rightarrow$ Colour Encoder} \label{sec:queryColourEncoder}

\begin{table*}[!ht]
\caption{Test losses for the models trained with different objective functions. The best results are highlighted in boldface.}
\label{colourResultsTesting}
\centering
\begin{tabular}{ C{1.6cm} C{1.4cm} C{1.4cm} c C{1cm} C{1cm} C{1cm} c C{1.2cm} }
    \toprule
    \multirow{2}{*}{\shortstack{\\Training\\Objective}} & \multirow{2}{*}{\shortstack{\\Training\\Loss}} & \multirow{2}{*}{\shortstack{\\Validation\phantom{g}\\Loss}} & \multicolumn{5}{c}{Test Metric} & \multirow{2}{*}{$\mathrm{D_{XKCD}}$}\\
    \cmidrule{5-7}
     & & & & $\mathrm{D_{KL}}$ & $\mathrm{D_{HI}}$ & $\mathrm{D_{LUV}}$ & &\\
    \midrule
    $\mathrm{D_{KL}}$     & 1.367 & 1.427 && \bf 1.445 & 0.616      & 0.018      && \bf 7.365 \\
    $\mathrm{D_{HI}}$     & 0.524 & 0.546 && 2.617     & \bf 0.564  & 0.028      && 10.588    \\
    $\mathrm{D_{LUV}}$    & 0.015 & 0.016 && 1.559     & 0.607      & \bf 0.016  && 8.055     \\
    \bottomrule
\end{tabular}
\end{table*}

Averages across clicked results enable us to represent queries from the historical query log in colour space. To be able to synthesize this embedding for unseen queries, we train a neural network that accepts a query as input and outputs the colour embedding. To this end, we construct a dataset that comprises of the query strings from our impression log, and their ground-truth colour representation computed as described in the previous section. To capture the context in a query, we use word embeddings to represent the words and pass this sequence through a bi-directional LSTM~\cite{graves2013generating} layer. The outputs of the LSTM layers are concatenated to form the \textit{query-features}. This query encoder is followed by a fully connected network and finally passed through a softmax layer that returns a distribution over the $327$ colour bins. 

\begin{figure}[!ht]
\centering
\includegraphics[width=0.65\linewidth]{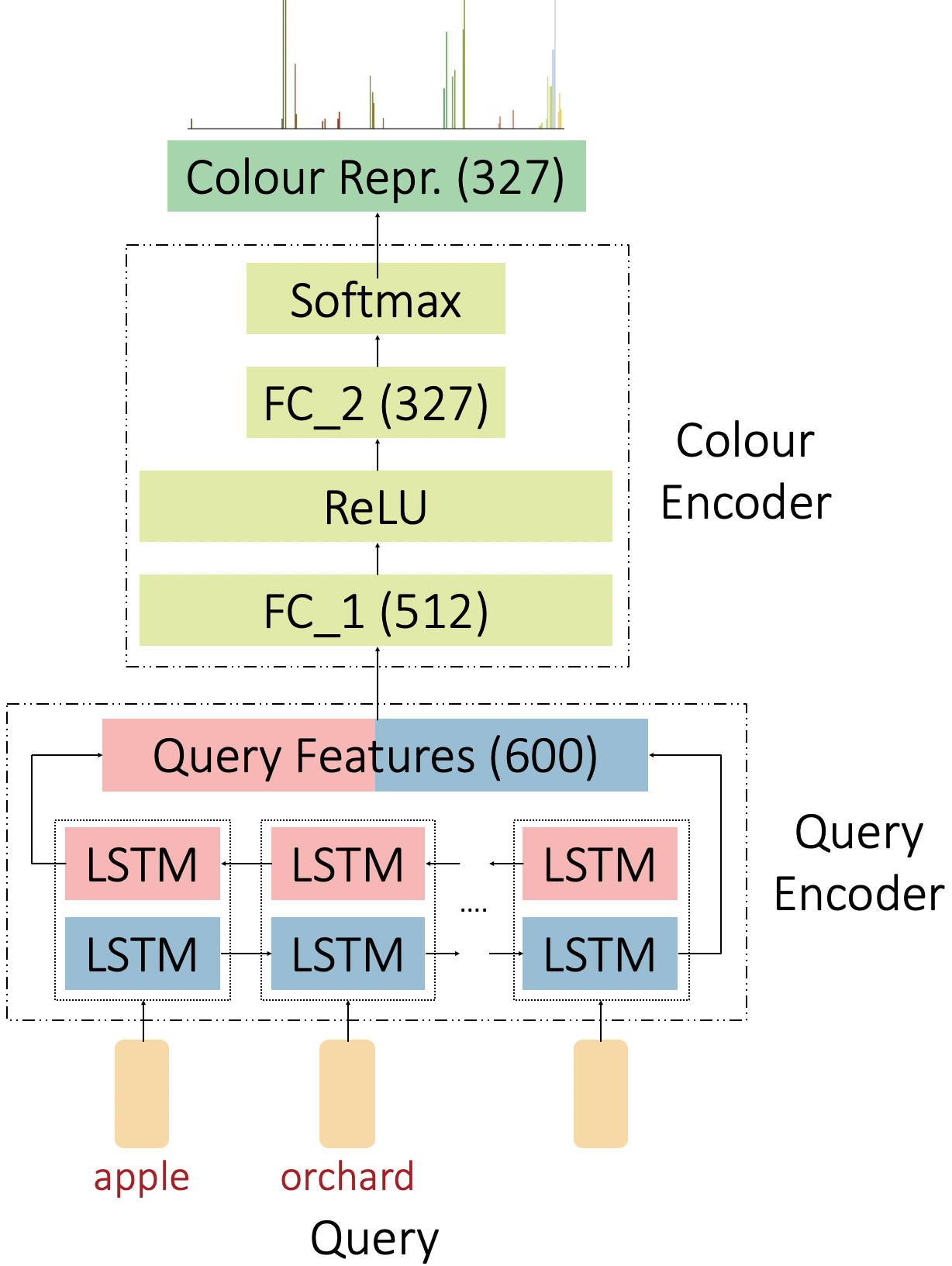}
\caption{Model architecture for the query $\rightarrow$ colour encoder. It predicts a colour embedding for a text query as a distribution over the quantised colour space.}
\label{model1}
\end{figure}

Recent progress on neural methods in IR has led to a variety of models to obtain distributed representations for queries -- e.g. \cite{xiong2017end, pang2017deeprank} deal with ad-hoc ranking, while the authors of \cite{HRED, mitra2015exploring} consider query suggestions and \cite{zhang2019} looks at intent classification. Our model architecture, summarised in Figure~\ref{model1}, is specialised for the focus on colour.

To guide the training of the model towards the desired target colour representation, we experiment with different objective functions that compute the distance between the predicted histogram $Q$ and target $P$. We use the distances described in Section~\ref{sec:colourSimilarity} as objective functions for this training.

\textbf{Implementation Details} : As a preprocessing step, we remove all non-alphanumeric characters (excluding spaces) in the user queries and convert all characters to lowercase. The query terms are embedded using $300$ dimensional GloVe~\cite{glove} vectors and the LSTM layers comprise of $300$ neurons each. The fully connected network after the query encoder consists of $2$ hidden layers of $1024$ and $512$ units respectively. The available data was split in the ratio $64:16:20$ for training, validation and testing. These sets were mutually exclusive, with no overlap amongst the queries in the three sets. We use Stochastic Gradient Descent optimizer with a batch size of $64$ and a learning rate of $0.01$ to train our model. The model is trained for $1000$ epochs and we report the results corresponding to the model with the lowest validation loss.

\begin{figure}[!ht]
\centering
\includegraphics[width=\linewidth]{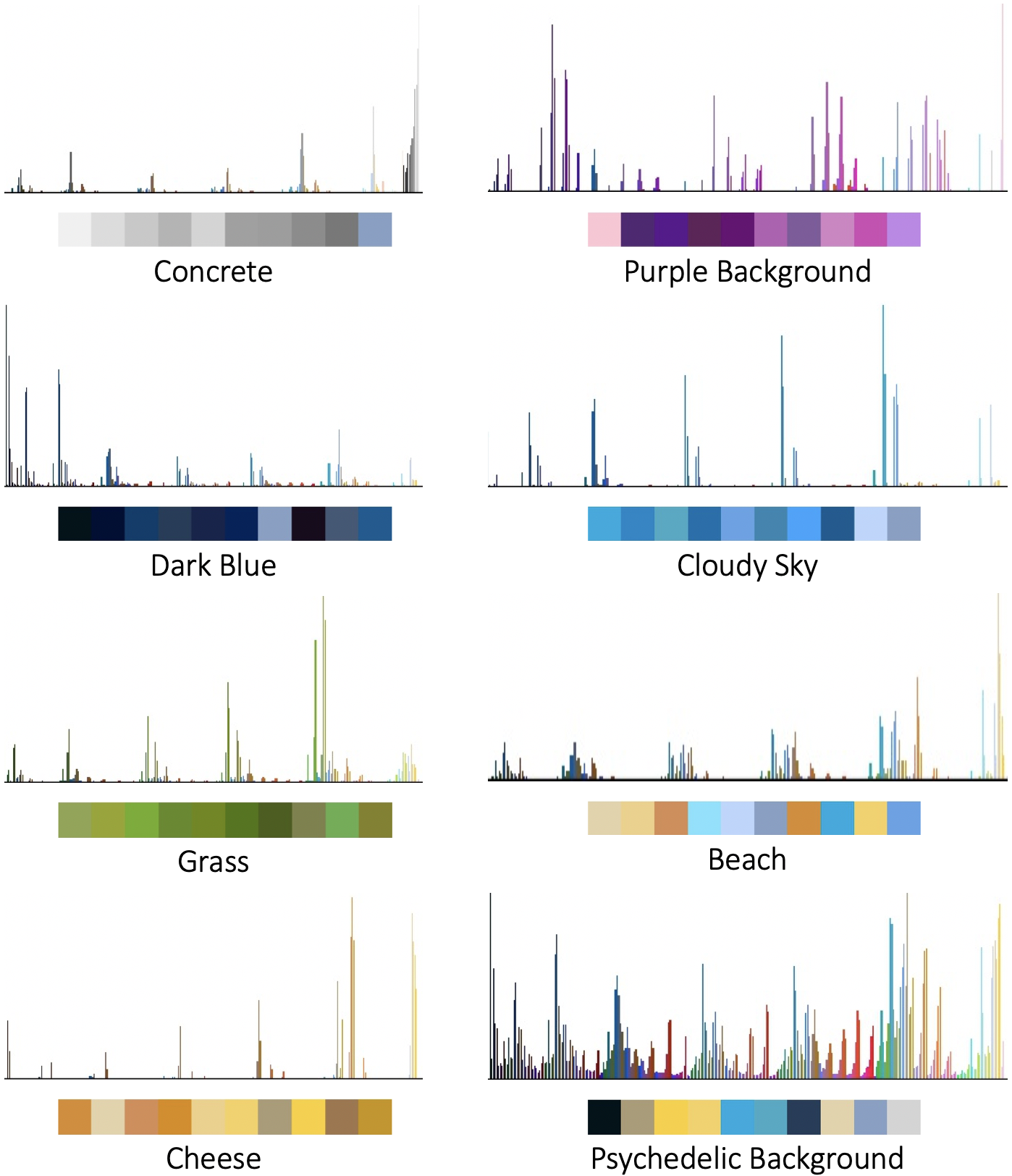}
\caption{Qualitative results for the query $\rightarrow$ colour encoder. Queries with predicted colour histograms and top 10 bins. Note that the top 10 bins of `Psychedelic Background' do not capture the expected colourfulness of the phrase which can be best appreciated in the complete histogram.}
\label{colourPreds}
\end{figure}

\textbf{Results} : The resulting performance (as measured by the corresponding objectives) of training on the previously mentioned metrics are shown in Table~\ref{colourResultsTesting}. The comparison across training and validation indicates that the learning process of our query $\rightarrow$ colour encoder is well behaved, i.e., over-fitting and similar concerns are handled well. In Figure~\ref{colourPreds}, we provide model predictions for select queries as qualitative evidence for the performance of the model.

Comparison across the different training objectives is however not straightforward. We construct a $3*3$ matrix where the entry in row $i$ and column $j$ represents the value of the $j^{th}$ test metric when the model was trained on the $i^{th}$ objective. Since the diagonal in the \textit{Test Metric} section of Table~\ref{colourResultsTesting} dominates, it indicates that the test loss is the smallest when the corresponding objective was used for training. While this behaviour is expected, it does not provide a conclusive choice for the more appropriate training objective for our task. It would be desirable for a particular training objective (row) to be consistently good across all test objectives (column).

Both KL-Divergence ($\mathrm{D_{KL}}$) and LUV Distance ($\mathrm{D_{LUV}}$) provide comparable performance. Note that $\mathrm{D_{KL}}$ is a generic distance metric between distributions with no specific notion of colour, while $\mathrm{D_{LUV}}$ is motivated by well-known studies in this domain. Histogram intersection ($\mathrm{D_{HI}}$) is another colour-agnostic training objective whose lower performance can be attributed to it being a strict metric which is hard to optimise.

\subsection{Evaluation on XKCD} \label{sec:XKCDTesting}
\begin{figure}[!ht]
\centering
\includegraphics[width=\linewidth]{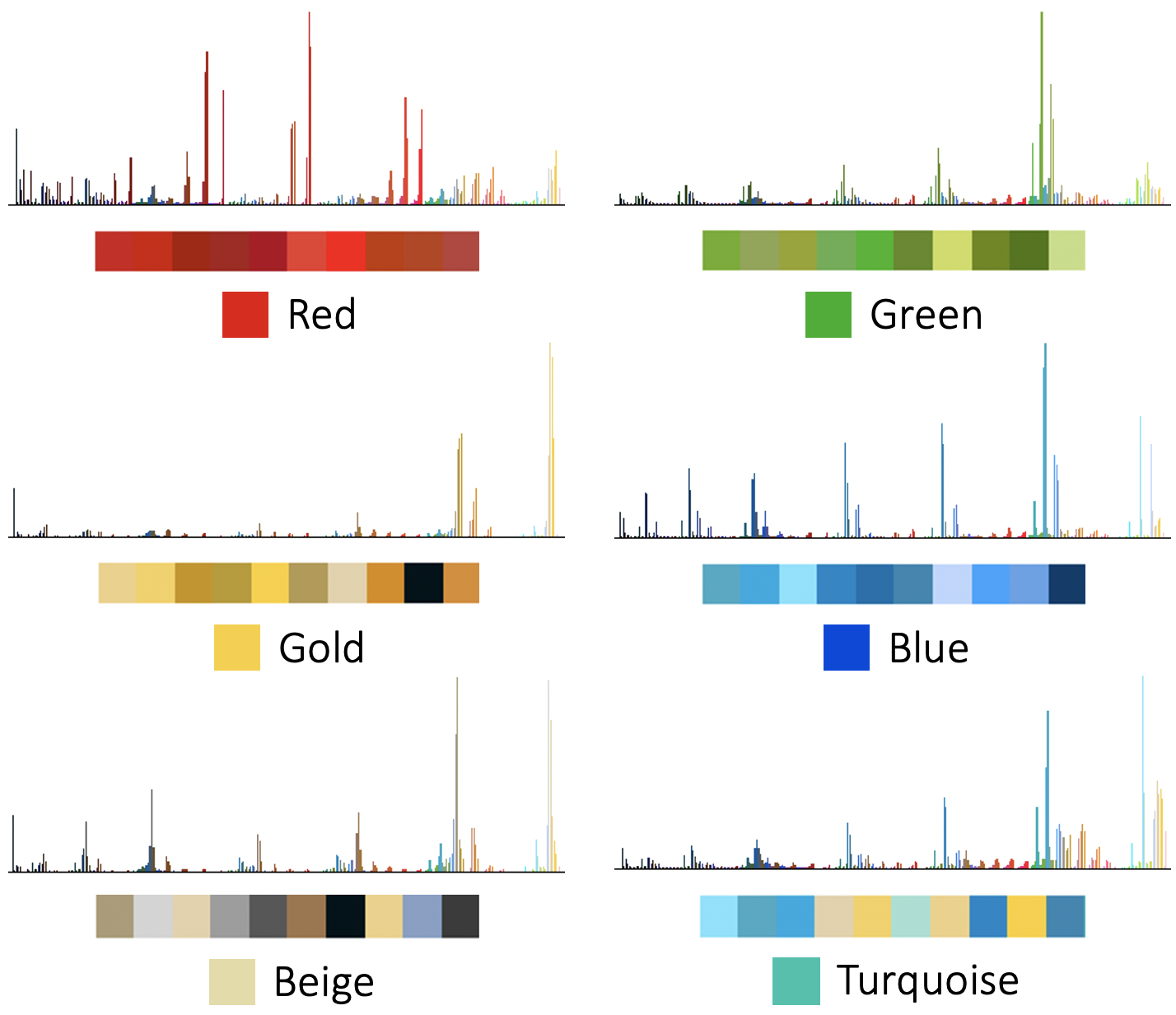}
\caption{Qualitative results on the XKCD dataset. Each subplot shows the colour name and RGB point in XKCD, and the predicted colour histogram and its top 10 bins.}
\label{xkcd_commonColors}
\end{figure}

To ascertain our model's ability to generalize, we test them on the standard XKCD dataset~\cite{xkcd}, a large crowd-sourced colour-naming survey. This dataset contains mappings for $\sim2.3$ Million colour names to points in RGB space. Since the dataset provides a mapping to a single RGB value, we need a metric to evaluate it against the output our model produces (a distribution over colour bins). We compute the negative log-likelihood of our output histograms given the XKCD labels in $P$ as
\begin{displaymath}
    \mathrm{D_{XKCD}} (P, Q) = - \sum_i^B \mathbb{1}_{(P_i\neq0)} * \log Q_i
\end{displaymath}
The ideal scenario when the model predicts the correct colour bin with probability $1$ yields a $\mathrm{D_{XKCD}}$ score of $0$. In all other cases, the metric strongly discourages the placement of any probability mass in other bins. It is evident from Table~\ref{colourResultsTesting} that our models perform well even on text phrases from the XKCD dataset, which largely do not resemble search queries. This is further substantiated by the qualitative results in Figure~\ref{xkcd_commonColors}, wherein the top $10$ bins of the histogram are highly correlated with the XKCD ground-truth colour. The proposed metric, $\mathrm{D_{XKCD}}$, also allows us to compare across different objective functions, with $\mathrm{D_{KL}}$ performing better than the other two functions.

\begin{figure}[!ht]
\centering
\includegraphics[width=\linewidth]{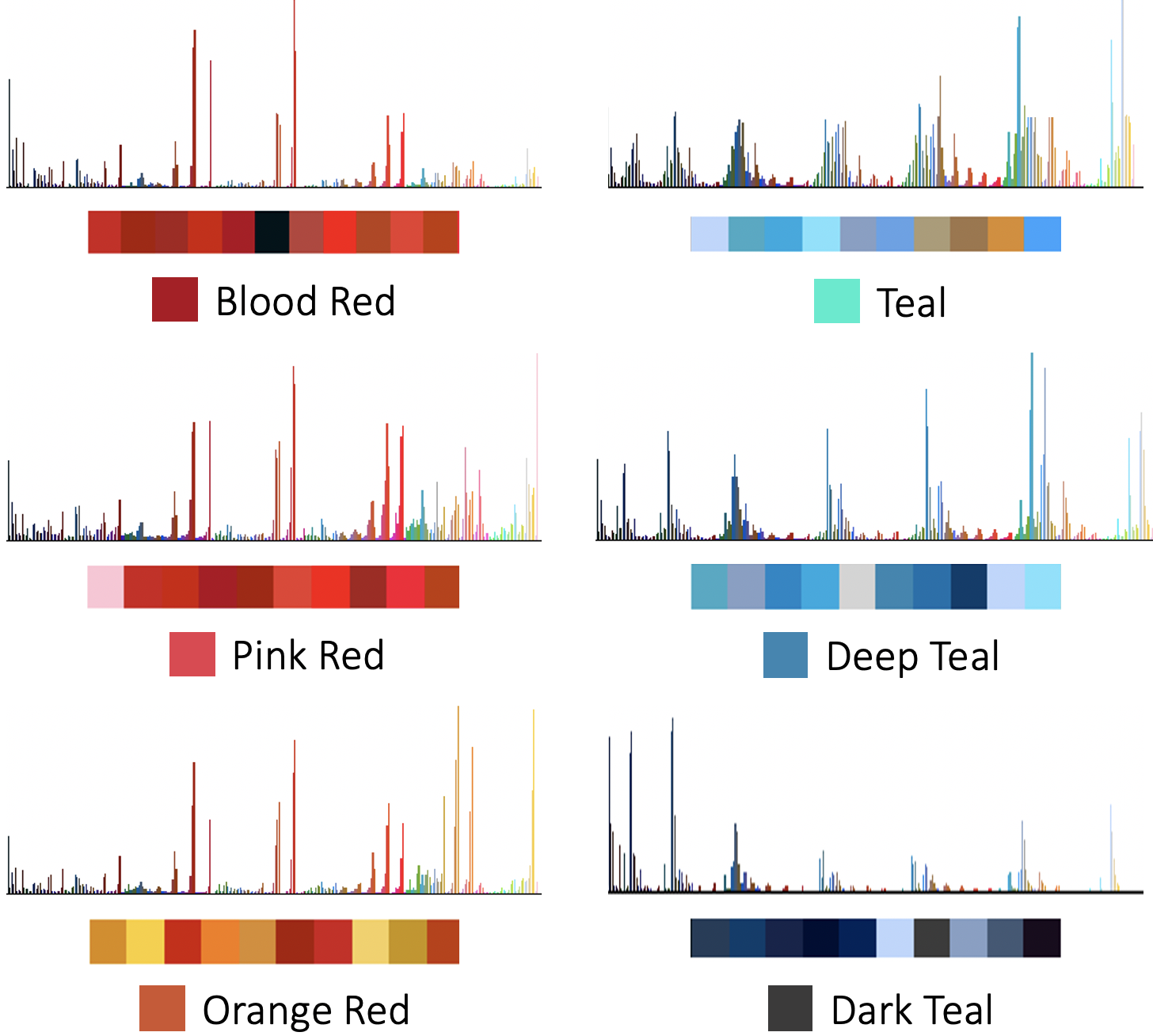}
\caption{Colour variations in the XKCD dataset. Each subplot shows the colour name and RGB point in XKCD, and the predicted colour histogram and its top 10 bins.}
\label{xkcd_variations}
\end{figure}

While Figure~\ref{xkcd_commonColors} indicates that the models are able to capture knowledge about specific colours, we would like to evaluate their behaviour on phrases that contain qualifiers coupled with colour words. As can be seen in Figure~\ref{xkcd_variations}, our model is also able to learn the subtle differences in colour representation that arise because of colour combinations (such as ``Pink Red'' and ``Orange Red'') or intensifiers (such as ``Deep'' and ``Dark'') or commonsense knowledge (such as ``Blood Red''). 

\subsection{Human Evaluation}
For a perceptual evaluation of the colour embeddings, we conducted an online survey using Amazon Mechanical Turk\footnote{https://www.mturk.com/} where participants were asked to rate the relevance between a textual phrase and a palette representing a colour histogram. For this purpose, we sampled at random $50$ queries from the test set of search queries (from the query logs) which were also present in the XKCD dataset. By restricting ourselves to the textual phrases within the XKCD set of named colours, we are attempting to ensure that the user study covers queries that have significant colour intent. While this set of chosen queries may not be representative of the search engine's query load (where not all queries contain colour terms), we believe that it is sufficiently rich in its composition (e.g. it contains commonplace objects like ``chocolate'', ``amethyst'' and ``dirt''). Restricting ourselves to a standardised dataset allows us to anchor the evaluation of the models that we have built for the novel task of text $\rightarrow$ colour-representation proposed in this paper. 

Alongside the query, users were shown either the ground-truth histogram (average of clicked images' colour representations) or the corresponding model output. We use the model trained on $\mathrm{D_{KL}}$ to make predictions as it outperformed the other two training objectives in Section~\ref{sec:XKCDTesting}. The Turkers were asked to rate their relevance on a $5-$point Likert scale ranging from \textit{Not at all relevant} to \textit{Extremely relevant}. Using such a scoring scale serves a twofold purpose - (1) to measure the quality of the proposed ground-truth colour representation through a crowd-sourced survey, and (2) to evaluate the performance of the model by comparing model prediction against ground-truth. Every \{\hspace{0.5ex}query, histogram\hspace{0.5ex}\} pair was annotated by $10$ different workers, and we confined ourselves to users with at least $100$ approved annotations and over $95\%$ acceptance rate. A total of $1000$ responses were collected, $500$ for each of ground-truth and predicted histograms.

\begin{table}[!ht]
\caption{Mean and standard deviation of ratings (scale 1-5) for ground-truth and predicted histograms.}
\label{humanEvaluation}
\centering
\begin{tabular}{C{2.3cm} C{2.5cm} }
    \toprule
    Histograms & Rating \\
    \midrule
    Ground-Truth    & $3.81\pm1.07$ \\
    Model Output & $3.49\pm1.18$\\
    \bottomrule
\end{tabular}
\end{table}

The findings of the human survey are summarized in Table~\ref{humanEvaluation}. To judge the significance of the results, we performed one-way ANOVA testing and the difference was found to be statistically significant with $p < 0.05$. We measured the inter-annotator agreement using Krippendorff's alpha coefficient~\cite{Krippen} and a score of $0.39$ indicates a positive agreement among the MTurk users. A mean rating of $3.81$ for the ground-truth colour representation indicates the correctness of the label-gathering method proposed. Majority of the users found the colour profile to be highly relevant to the given query and this corroborates our intuitions about incorporating human feedback. The average rating for ground-truth histograms was slightly higher than model outputs, which is expected -- this difference is tolerable because the end-goal of this learnt embedding is to contribute to the ranking function of a retrieval system, rather than as part of a user-facing interface.

\section{Colour as a Ranking Feature} \label{seq:colourFtr}
The models described in the previous section were trained on data collected from a search setting. We hypothesise that a notion of colour can help improve the ranking quality in an image retrieval application. In this section, we train a cross-modal network that learns to rank images for a given textual query. We follow standard practice~\cite{ngiam2011multimodal, wang2016learning, wang2018learning} while designing the architecture of this network. We use initial modality-specific layers that produce embeddings independently for the two modalities (images and associated metadata, and the textual query) followed by fusion layers that combine the information from the two sub-networks. 

We consider \textit{query features} obtained from word-level embeddings passed through an LSTM. On the image side, we have the following metadata - image content, author provided captions (sentence-length textual descriptions) and tags (representative of the salient objects in the image). We embed the images using a ResNet~\cite{he2016deep} model, which is the state-of-the-art network in various computer vision tasks including image classification. We fine-tune the last $2$ layers of the pre-trained model on our image collection. These image embeddings, along with those for the captions and tags (obtained by individually taking the average of the corresponding word embeddings) provide our list of \textit{image features}. These are concatenated to produce the final feature representation of an image.

This network is trained using the boolean clicked-or-not labels available from query logs and learns to predict a relevance score ($0$ to $1$) for a \{\hspace{0.5ex}query, image\hspace{0.5ex}\} pair. Our baseline model uses the features as described above and is trained on the RankNet objective~\cite{Burges2005}:
\begin{equation} \label{eq:rankLoss}
    \mathcal{L_R} = - \frac{1}{m^2} \sum_{j=1}^m\sum_{ \substack{k=1\\k\neq j} }^m \bigg(\, y_{jk}*\log \hat{y}_{jk}  + (1 - y_{jk})*\log(1 - \hat{y}_{jk})\, \bigg)
\end{equation}
where, $\hat{y}_{jk} = p(s_i^j > s_i^k) = \sigma(s_i^j - s_i^k)$ indicates the probability of result $j$ being ranked higher than result $k$. The $y_{jk}$ are obtained from the original click data by setting $y_{jk}=1$ if the result $j$ was clicked and $k$ was not, and $\sigma(x)$ is the sigmoid function. The cross-entropy loss is summed over all pairs of images for a given query query, with $m$ denoting the number of associated images. $s_i$ and $s_j$ are the output scores of the model, and the model parameters are learnt by optimising for $\mathcal{L_R}$. 

While we have utilised RankNet as the training objective for the cross-modal ranking function~\cite{wang2016comprehensive}, other options are possible. For example, simply building a clicked-vs-not classifier~\cite{lu2014learning}, or optimising any of the other standard learning-to-rank objectives~\cite{liu2009learning}. But our primary objective is to evaluate the potential gains in ranking quality due to a colour-centric feature, given an image search ranker. To establish this, we keep all aspects of the \textit{baseline} (model architecture, training method, query and image features) constant with the addition of colour information being the only change.

We experiment with the addition of colour information to the baseline model in the following two variations:
\begin{enumerate}
    \item \textbf{Baseline + Colour Representation} : In addition to the aforementioned features, the colour vectors for both query and image are input to the model. The image histogram is added to the \textit{image features} and the colour representation of query to \textit{query features}. This means that $2*327$ additional features are now passed as input to the fusion layers.
    \item \textbf{Baseline + Colour Distance} : Instead of adding the histograms of both modalities independently, we measure the distance between the two histograms for every \{\hspace{0.5ex}query, image\hspace{0.5ex}\} pair and append this value to the \textit{query-image features} of the model. The distances are computed using the functions described in Section~\ref{sec:colourSimilarity}. In this case, the model only has $1$ additional feature when compared to the baseline set.
\end{enumerate}
Note that in this section, we evaluate the utility of the ground-truth colour representation of queries as a ranking feature. We evaluate the performance of the query $\rightarrow$ colour encoder and learnt query representations in Section~\ref{multitask}.\\

\textbf{Implementation Details} : All word embeddings (for query terms, image tags and captions) are computed using $300$ dimension GloVe~\cite{glove} vectors. The LSTM layer consists of $300$ neurons and the hidden state of the final word is used to represent the query. Images are passed through a pre-trained ResNet network and the $2048-$dimension output of the last fully connected layer is taken as the image embedding. For the model variations, the colour representation of both modalities are histograms of length $327$ and colour distance is a scalar value. The set of \textit{query features} and \textit{image features} are concatenated to form the \textit{query-image features} and passed through a fully connected network with ReLU activation to produce the relevance score. We use the same dataset split, learning rate and number of epochs as in the query $\rightarrow$ colour encoder. Additionally, back-propagation using Stochastic Gradient Descent is performed after every query (i.e. batch size $1$) given the formulation of the RankNet objective. The clicked images for a query are positive examples, and for the corresponding negative example -- we retain a skipped (not clicked) image with probability $0.9$ and sample a random image from the entire collection with the remaining probability. Statistical significance of the obtained metrics is verified using a Paired \textit{t}-test.\\

\begin{table}[H]
\caption{Ranking performance for different models. The best results are highlighted in boldface. ${ }^\dagger$ indicates that the result are statistically significant at 0.05 level.}
\label{rankingResults}
\centering
\begin{tabular}{C{3.3cm} C{1cm} C{1cm} C{1cm}}
    \toprule
     & AUC & MAP & MRR\\
    \midrule
    Baseline                        & $0.662$      & $0.264$      & $0.406$     \\[0.6ex]
    \hdashline\noalign{\vskip 0.6ex}
    Baseline + Colour Repr.         & $0.670^\dagger$      & $0.270^\dagger$      & $0.417^\dagger$     \\
    Baseline + $\mathrm{D_{KL}}$    & $\bf{0.744}^\dagger$  & $\bf{0.430}^\dagger$  & $\bf{0.647}^\dagger$ \\
    Baseline + $\mathrm{D_{HI}}$    & $0.713^\dagger$      & $0.370^\dagger$      & $0.578^\dagger$     \\
    Baseline + $\mathrm{D_{LUV}}$   & $0.679^\dagger$      & $0.286^\dagger$      & $0.453^\dagger$     \\
    \bottomrule
\end{tabular}
\end{table}

\textbf{Results} : We measure the performance of the models using ranking metrics - Area Under the Curve (AUC), Mean Average Precision (MAP) and Mean Reciprocal Rank (MRR). These are computed over the clicked images of the test set and the results are reported in Table~\ref{rankingResults}. It can be seen that all the model variants with colour information perform better than the \textit{Baseline} model. This validates our hypothesis about colour being an important feature that guides ranking, and hence, human discernment. This is further corroborated by the significant improvement in metrics for \textit{Baseline}$ + \mathrm{D_{KL}}$ with just one additional feature in the input. Moreover, $\mathrm{D_{KL}}$ consistently outperforms the other two distance functions in this ranking setup.

\section{Intermediate Layer Regularisation} \label{multitask}

\begin{table*}[!ht]
\caption{Jointly learning the ranker and colour representation: Evaluation on the query $\rightarrow$ colour task. These results can be compared to Table~\ref{colourResultsTesting} where the query $\rightarrow$ colour encoder was trained in isolation. $\ast$ indicates that the metric outperforms the corresponding value in Table~\ref{colourResultsTesting}. The best results are highlighted in boldface.}
\label{multiTaskResults_colour}
\centering
\begin{tabular}{ C{2.0cm} C{1.4cm} C{1.4cm} c C{1cm} C{1cm} C{1cm} c C{1.2cm} }
    \toprule
    \multirow{2}{*}{\shortstack{\\Training\\Objective}} & \multirow{2}{*}{\shortstack{\\Training\\Loss}} & \multirow{2}{*}{\shortstack{\\Validation\phantom{g}\\Loss}} & \multicolumn{5}{c}{Test Metric} & \multirow{2}{*}{$\mathrm{D_{XKCD}}$}\\
    \cmidrule{5-7}
     & & & & $\mathrm{D_{KL}}$ & $\mathrm{D_{HI}}$ & $\mathrm{D_{LUV}}$ & & \\
    \midrule
    $\mathrm{\hat{D}_{KL}}\,(\alpha = 0.5)$ & $1.225^\ast$ & $1.277^\ast$ & & $\bf{1.282}^\ast$ & $\bf{0.566}^\ast$ & $\bf{0.014}^\ast$ & & $6.291^\ast$ \\
    $\mathrm{\hat{D}_{HI}}\,(\alpha = 1)$ & $0.658$ & $0.663$ & & $4.283$ & $0.662$ & $0.149$ & & $10.289^\ast$ \\
    $\mathrm{\hat{D}_{LUV}}\,(\alpha = 4)$ & $0.065$ & $0.066$ & & $3.01$ & $0.866$ & $0.064$ & & $\bf{5.794}^\ast$ \\
    \bottomrule
\end{tabular}
\end{table*}

In the previous section, we examined the role that colour might play in improving search ranking. We did this by providing as input the ground-truth colour representation of a query (obtained as an average of colour histograms of clicked images) to a baseline cross-modal ranker. Given the experience with models to learn a query's colour representation, we extend this setup to simultaneously optimise the cross-modal ranker and the query $\rightarrow$ colour encoder in an end-to-end manner. This is achieved by regularising an intermediate layer on the \textit{query side} model to represent the colour information of the query. 

\begin{figure}[!ht]
\centering
\includegraphics[width=\linewidth]{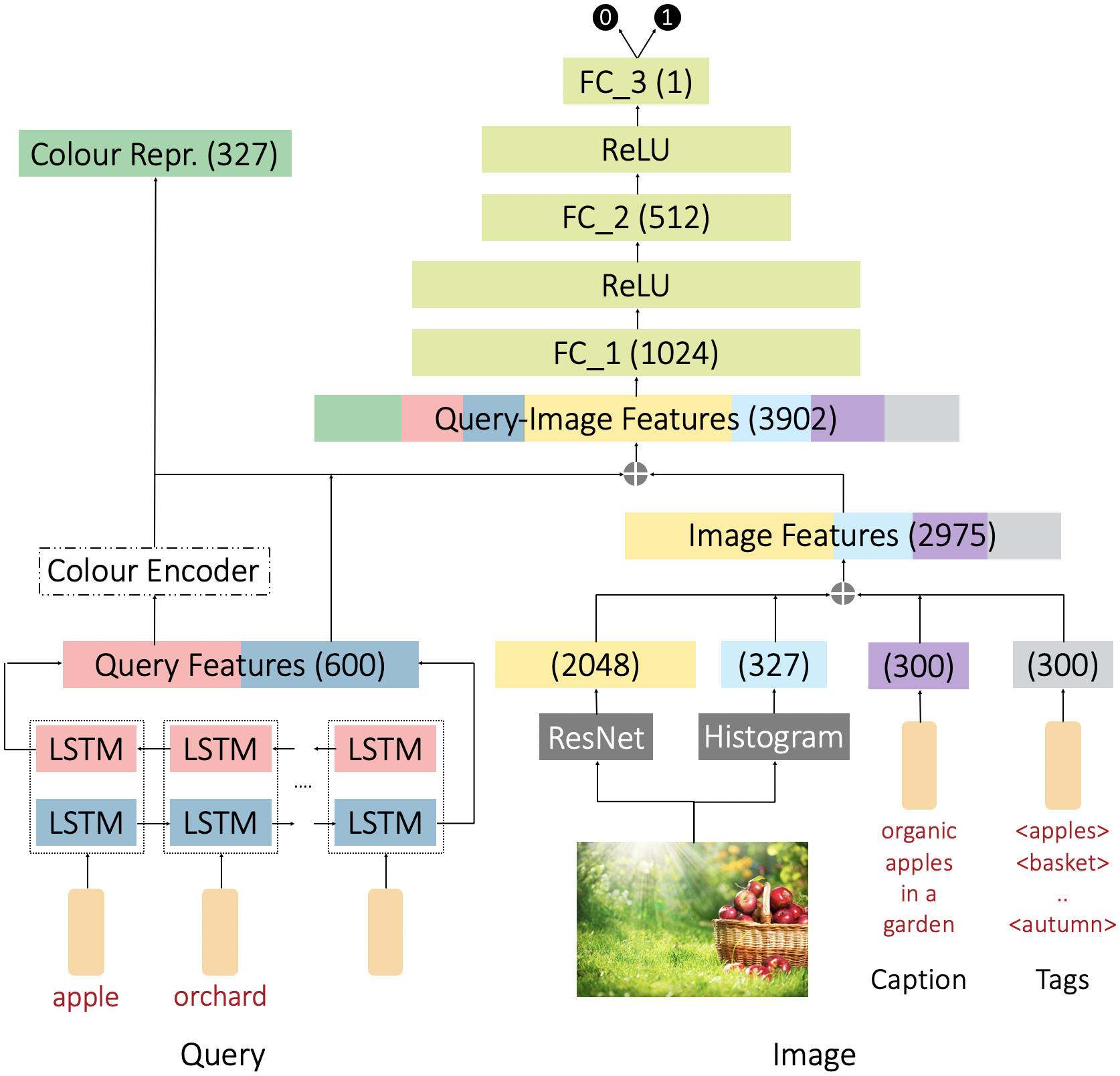}
\caption{Model Architecture for intermediate layer reqularisation: joinly learning the search ranker and colour representation for text queries.}
\label{model2}
\end{figure}

The model architecture, summarised in Figure~\ref{model2}, is a combination of the query $\rightarrow$ colour encoder and the search ranker. As in the previous section, \textit{image} and \textit{query features} are individually passed through modality-specific layers before being combined. In contrast to the previous section, the colour information for the query is the output of the query $\rightarrow$ colour encoder, rather than the ground-truth. The architecture of this module is retained from that described in Section~\ref{sec:queryColourEncoder}. To exploit the synergy between the two tasks, we allow parameter sharing between the two models by having a shared representation in the query encoder.

In this model, the output of the query $\rightarrow$ colour encoder is an internal $327$-dimension vector. To semantically ground it in the notion of colour, we introduce a term in the training objective that captures how close this embedding is to the ground-truth colour representation for this query. Therefore, the final training objective comprises of two different components -- RankNet loss and query-colour encoding loss. The combined objective function can be expressed mathematically as,
\begin{displaymath}
    \mathcal{L_R} + \frac{1}{1+\alpha}\,\mathrm{D_{*}} (P, Q)
\end{displaymath}
where $\mathcal{L_R}$ is the RankNet loss (Equation~\ref{eq:rankLoss}), and as before, $P$ is the predicted colour histogram while $Q$ is the corresponding label. Here, $\mathcal{L_R}$ is defined at a query level and $\mathrm{D_{*}}(\cdot)$ is averaged over all images of that query. $\alpha (> 0)$ is the regularization hyper-parameter that controls the relative contribution of the two losses. It is tuned for different colour losses using the validation set. For each loss function $\mathrm{D_{*}}(\cdot)$, the optimal value of $\alpha$ is chosen using MAP as the validation metric. 

\begin{table}[!ht]
\caption{Jointly learning the ranker and colour representation: Evaluation on the clicked-vs-not task. The best results are highlighted in boldface. ${ }^\dagger$ indicates that the result is statistically significant at 0.05 level.}
\label{multiTaskResults_ranking}
\centering
\begin{tabular}{C{2.3cm} C{1.1cm} C{1.1cm} C{1.1cm}}
    \toprule
    & AUC & MAP & MRR\\
    \midrule
    Baseline    & $0.662$     & $0.264$     & $0.406$ \\[0.6ex]
    \hdashline\noalign{\vskip 0.6ex}
    $\mathrm{\hat{D}_{KL}}\,(\alpha = 0.5)$   & $0.663^\dagger$     & $0.272^\dagger$     & 0$.427^\dagger$ \\
    $\mathrm{\hat{D}_{HI}}\,(\alpha = 1)$     & $0.669^\dagger$     & $\bf{0.274}^\dagger$ & $\bf{0.435}^\dagger$ \\
    $\mathrm{\hat{D}_{LUV}}\,(\alpha = 4)$    & $\bf{0.673}^\dagger$ & $0.268^\dagger$     & $0.418^\dagger$ \\
    \bottomrule
\end{tabular}
\end{table}

\textbf{Results} : We report the metrics for evaluating the model predictions in Table~\ref{multiTaskResults_colour} (query $\rightarrow$ colour task) and Table~\ref{multiTaskResults_ranking} (clicked-vs-not ranking task). We have used the notation $\mathrm{\hat{D}}$ to indicate that distances were computed based on the output of the query $\rightarrow$ colour encoder, differentiating from $D$ in Table~\ref{rankingResults} which used the ground-truth. It is evident that the values of each of the ranking metrics in Table~\ref{multiTaskResults_ranking} is marginally lower than the corresponding values in Table~\ref{rankingResults}. This can be attributed to the difference between using the ground-truth versus utilising a model prediction. Even then, for all choices of colour-distance metric $\mathrm{D_{*}}(\cdot)$, the achieved MAP and MRR in Table~\ref{multiTaskResults_ranking} are significantly better than the baseline values. Table~\ref{multiTaskResults_colour} contains the evaluation of the colour-representation prediction task. The multi-task setup benefits the training for $\mathrm{D_{KL}}$, evident from the lower values when compared to Table~\ref{colourResultsTesting}. This implies that the additional information content from the ranker network further guides the learning of the colour representation. The same is not true for $\mathrm{D_{HI}}$ and $\mathrm{D_{LUV}}$ and this requires further investigation into colour spaces and distances.

\section{Conclusion}
In this paper, we have considered the novel task of learning a colour representation of search queries, i.e, a neural network that maps a given string into colour space. While there is prior work that considers the association in both directions -- colour $\leftrightarrow$ textual-name -- these have typically been with controlled vocabularies. We have removed restrictions on both sides: (a) rather than a predefined set of colour words, we would like our models to work on the full range of long-tail words used in search queries, and (b) we consider a distribution over colour space rather than a single point. 

Our experimental results show that a query $\rightarrow$ colour encoding model trained on clicked data allows the addition of a \textit{colour match} feature into a search ranker that leads to an increase in retrieval metrics like MAP and MRR. A crowd-sourced human survey and thorough qualitative analysis indicate the viability of our colour representations and the learning ability of proposed models. In addition to these, we believe that the model described here has two advantages: (1) the output of the model is an interpretable representation that can be examined visually, and (2) once trained, the model can be utilized in other language and vision tasks where a text $\rightarrow$ colour mapping is desired.

While the experiments show positive results for the building of these models, there are many avenues for future work. On the query side, richer representations of queries~\cite{zhang2019} is a natural area for exploration. In addition, since colour words in queries are typically used as adjectives of objects, being able to reason separately about the different components of the query~\cite{Misra_2017_CVPR} in a scalable manner might be beneficial. In our work, we have worked with a given choice of colour space and representation for the images. Alternative colour spaces have different relative benefits and strengths, especially in how they reflect human perception. A thorough treatment of all these aspects is required to enable our goal of building data-driven models of colour.



\bibliographystyle{ACM-Reference-Format}
\bibliography{main}

\end{document}